\documentclass[journal]{IEEEtran}

\usepackage{setspace}
\usepackage{lettrine}
\usepackage{tikz}
\usepackage{graphics,
	psfrag,
	epsfig,
	amsthm,
	cite,
	amssymb,
	url,
	dsfont,
	algorithmic,
	balance,
	enumerate,
	color,
	setspace,
    soul
}
\usepackage{amsmath}
\usepackage{tabularx}
\usepackage{epstopdf}
\usepackage{mathtools}

\usepackage{footmisc}

\usepackage{optidef}

\newcommand{\cref}[1]{Constraint~\ref{#1}}

\newcommand{\ignore}[1]{}

\usepackage[labelformat=simple]{subcaption}

\IEEEoverridecommandlockouts
\usepackage{mathtools}
\usepackage{color}
\usepackage[utf8]{inputenc}
\usepackage[ruled, lined, longend]{algorithm2e}
\SetEndCharOfAlgoLine{}

\begin{document}

		\IEEEoverridecommandlockouts

\title{Smart Helper-Aided  F-RANs: \\Improving  Delay and Reducing Fronthaul Load}

\author{\IEEEauthorblockN{Hesameddin~Mokhtarzadeh, \IEEEmembership{Student~Member,~IEEE,} Mohammed~S.~Al-Abiad, ~\IEEEmembership{Member,~IEEE,} \\Md Jahangir Hossain, \IEEEmembership{Senior~Member,~IEEE,} Julian Cheng, \IEEEmembership{Fellow,~IEEE} }\\
\IEEEcompsocitemizethanks{\IEEEcompsocthanksitem Hesameddin~Mokhtarzadeh, M. J. Hossain, and Julian Cheng are with the School of Engineering, The University of British Columbia, Kelowna, BC V1V 1V7, Canada (e-mails: hesameddin.mokhtarzadeh@ubc.ca, jahangir.hossain@ubc.ca, julian.cheng@ubc.ca).

Mohammed S. Al-Abiad is with the Department of Electrical and Computer Engineering, University of Toronto, Toronto, ON M5S, Canada (e-mail: mohammed.saif@utoronto.ca).}	}

 \maketitle
 \date{}

\begin{abstract}
In traditional Fog-Radio Access Networks (F-RANs), enhanced remote radio heads (eRRHs) are connected to a macro base station (MBS) through fronthaul links. Deploying a massive number of eRRHs is not always feasible due to site constraints and the cost of fronthaul links. This paper introduces an innovative concept of using smart helpers (SHs)  in  F-RANs. These SHs do not require fronthaul links and listen to the nearby eRRHs' communications. Then, they smartly select and cache popular content. This capability enables SHs to serve users with frequent on-demand service requests potentially. As such, network operators have the flexibility to easily deploy SHs in various scenarios, such as dense urban areas and temporary public events, to expand their F-RANs and improve the quality of service (QoS). To study the performance of the proposed SH-aided F-RAN, we formulate an optimization problem of minimizing the average transmission delay that jointly optimizes cache resources and user scheduling. To tackle the formulated problem, we develop an innovative multi-stage algorithm that uses a reinforcement learning (RL) framework. 
Various performance measures, e.g., the average transmission delay, fronthaul load, and cache hit rate of the proposed SH-aided F-RAN are evaluated numerically and compared with those of traditional F-RANs. 

\end{abstract}
\begin{IEEEkeywords}
	
F-RANs, low latency communications, reinforcement learning, resource allocation, smart helpers. 
\end{IEEEkeywords}

\section{Introduction}
\IEEEPARstart{I}{n} the digital era, there is a proliferation of connected devices and services that are to be supported by wireless networks  
\cite{1, VR1, VR3}. As a result, there is a pressing need for wireless networks that can provide superior performances, such as ultra-reliable low latency communications (URRLC). 
To address this need, network densification by expanding the network access layer with more remote radio heads (RRHs) is shown to be a promising solution. 
Cloud-radio access networks (C-RANs) were proposed by China Mobile \cite{chinamobile} as a potential network architecture that supports network densification. In C-RANs, the baseband units (BBUs) of the RRHs are centralized within the macro base station (MBS), also called the central unit (CU). The RRHs are then connected to the MBS via the fronthaul links \cite{9152159}.

The prime concern of centralized load within the CU is that it leads to notable transmission delay and heavy traffic on fronthaul links.
More recently, Fog-RANs (F-RANs) have been introduced to address these issues \cite{7558153}. 
In F-RANs, the conventional RRHs are replaced with enhanced RRHs (eRRHs) with more sophisticated signal processing capabilities. More importantly, eRRHs have caching capabilities. 
F-RANs offer two key benefits as mentioned as follows.
First, network functionalities can be divided between eRRHs and the cloud based on the specific latency and reliability requirements. Furthermore, it enhances the efficiency and speed of content delivery as the eRRHs can serve the cached content to the users without frequently requesting these contents from the MBS \cite{FRANInt, FRANInt1}. Consequently, caching capability helps to reduce the load on the fronthaul link by reducing the amount of data that needs to be retrieved from the MBS \cite{FRANI1, FRANI2}. Network densification, by incorporating more eRRHs, offers improved file caching and delivery capabilities. Nevertheless, this enhancement comes at the expense of significantly increased hardware and energy consumption. In addition, deploying a large number of eRRHs can be challenging in densely populated urban areas, as it may be constrained by site limitations and the scarcity of available space. Even if a massive deployment of eRRHs is possible, there remains a significant challenge in managing the additional fronthaul link setup.

To further reduce the traffic load on fronthaul links that the centralized-based F-RAN architectures may face, cache-enabled device-to-device (CE-D2D) and cache-enabled relay (CE-relay) communication systems have been proposed \cite{D2D,CERelay1}. In CE-D2D communication systems, devices can cache popular content and assist in offloading eRRHs by transferring data between each other \cite{Cach}. In particular, eRRHs and CE-D2D users transmit their cached content to interested users using cellular and D2D links, respectively. 
Despite the potential benefits of CE-D2D communications, they did not gain popularity for various reasons, including privacy and power consumption concerns of devices. 
Additionally, CE-relays have been introduced to cache content and act as intermediaries between access points and users, delivering cached data to users during subsequent time frames. CE-relays cache desired content only after they have served a user with the same content previously. Hence, an occasional utilization of CE-rleays as a relay in the network causes inefficient cache resource updates, as well as improper communication history. Furthermore, frequent use of CE-relays as a relay in the network introduces extra delays and fronthaul loads due to the communication delays between eRRHs and CE-relays and the back-and-forth communications to the MBS. Therefore, establishing a balance between effective optimization of CE-relay cache resources and mitigating fronthaul load in multi-hop communications scenarios is of great importance which still remains a formidable challenge.



Optimized cache resources play a crucial role in efficiently serving users directly at the network edge \cite{8736961, FRANI7,FRANI3}. One trivial strategy is to predict the popularity of content requested by users, which helps in making decision in terms of   placing  various contents within the network \cite{FRANI7}. However, despite significant advancements in the prediction of content popularity, the dynamic nature of wireless networks, user mobility, and data access behavior necessitate a joint optimization of edge caching and wireless resource allocation \cite{FRANInt2, FRANInt3, mobility1, mobility2}. Achieving such joint optimization often requires exhaustive search methods due to the interdependence between caching decisions and resource allocation, which are not feasible for implementations \cite{Cache_Strat2, FRANI4}. 
To overcome the limitations of traditional optimization-based resource allocation algorithms, distributed machine learning (ML) algorithms have been proposed as potential solutions  \cite{FRAN3}. For example, reinforcement learning (RL) frameworks have been extensively explored to optimize caching decisions in an online manner, without prior knowledge of resource allocation decisions. These RL frameworks learn the access behavior of users and the mobility of the network, enabling the joint optimization of caching decisions and resource allocation within a distributed F-RAN architecture.
\subsection{Related Works} 
A flourish of works has been done towards maximizing the overall throughput on F-RANs, thereby improving the end-to-end latency. 
Existing works mainly focused on either optimizing radio resources, such as user scheduling and channel coding  \cite{RW1,RW2,RW7,RW4,RW5,RW8,RW6}, or integration of  caching with radio resource optimization strategies \cite{RW3, FRANI13,FRANI11,FRANI12,Game,FRANI8}. In what follows, we review the related works on radio resource optimization and joint caching and resource optimization. 
\subsubsection{Radio Resource Optimization} In \cite{RW1,RW2}, the authors proposed optimization strategies such as channel precoding, fronthaul compression, and superposition coding to reduce transmission latency and enhance the overall quality of service (QoS) in F-RANs. 
While these works have improved the performance of the network, they did not fully leverage the capabilities offered by the architecture of F-RANs. In addition, some research works have explored the benefits of cooperation between eRRHs in F-RANs.
For example, the work in \cite{RW7} proposed a distributed computing and content-sharing mechanism to enhance the file delivery process.
Although their collaborative system improves the QoS, it can also result in higher traffic on the links connecting these eRRHs together. 

To address this issue, the works in \cite{RW4, RW5} explored the capabilities of D2D communication modes. They demonstrated that incorporating D2D communication in F-RANs has the potential to ease the transmission burden on the eRRHs and significantly decrease the bandwidth usage of both RANs and fronthaul links. However, D2D communication may pose unexamined challenges, such as Asecurity risks and increased power consumption for the users. 
In addition, some recent research works investigated the potential improvement of delivery delay and outage probability by utilizing CE-relays \cite{CERelay2,CERelay3}. Specifically, in \cite{CERelay2}, the authors explored the benefits of cache-aided broadcast-relay wireless networks, demonstrating optimal tradeoffs between cache storage and latency, which is relevant to optimizing content delivery and network efficiency similar to CE-relays. Moreover, \cite{CERelay3} introduces a proactive caching strategy in CE-relay networks to address dynamically changing content request preferences environments. Nonetheless, the inherent problem of slow cache resource updates in CE-relays remains challenging as their capability to cache desired content depends on prior user requests.

Moreover, to enhance the transmission mechanism, several research studies have focused on investigating the placement of eRRHs and addressing the challenges associated with the dynamic nature of mobile eRRH network topology \cite{RW8,RW6}.
In \cite{RW8}, a study focused on maximizing throughput by dynamically determining optimal fog node locations using a clustering algorithm. Furthermore, an adaptive radio resource balancing scheme was proposed in \cite{RW6} to enhance the QoS in a system with mobile eRRHs. However, their study did not address the challenges associated with deploying high-capacity fronthaul links between mobile eRRHs and the CU. 

\subsubsection{Joint Cache and Resource Optimization} In \cite{RW3}, the authors performed an information-theoretic analysis to assess delivery latency relative to system interference. Their study shed light on the potential benefits of jointly optimizing caching and resource allocation on data delivery efficiency within this network architecture. In \cite{JointCandR1}, the authors addressed an optimization problem concerning cache and resource allocation to enhance energy efficiency in a heterogeneous Fog-RAN. The original problem, which is non-convex, is converted into an approximate convex format. While this transformation simplifies finding a solution, it may not precisely match the optimal solution of the initial non-convex problem due to the approximation involved. Similarly, in \cite{JointCandR2}, a joint resource allocation and caching placement problem in F-RANs was formulated. However, their approach assumes simultaneous optimization of cache and resource allocation, a method that often proves impractical, given that cache resources typically require updates over longer periods. The interdependence between caching decisions and resource allocation makes such joint optimization a complex task that can only be optimally solved via exhaustive search solutions,  which are not feasible in real-world applications.
In this regard, researchers have recently taken advantage of RL algorithms to tackle the challenges of such problems in F-RANs. For example, the authors of \cite{FRANI12} presented an RL framework on double deep Q-network for caching optimization and power allocation incorporating personalized user request modelling. Similarly, in \cite{FRANI13}, a deep RL-based algorithm was developed for joint dynamic cache resource optimization and power allocation to minimize the total transmission cost.
Additionally, in \cite{FRANI11}, an optimization problem was presented to minimize the weighted system cost for the uplink input data transmission, specifically to access the cached requested services of the fog access points. The authors adopted a two-timescale approach, combining a game-based resource allocation on a small timescale with a multi-agent RL (MARL) algorithm for caching decisions on a larger timescale. However, it should be noted that the problem of optimal user assignment was explicitly not addressed in these research works.
In \cite{Game}, a joint RL algorithm and game-theoretic approach for cache optimization and user assignment was proposed to maximize the sum data rate of the network, considering fixed transmission power for each fog node.
In \cite{FRANI8}, a more comprehensive system model was introduced, incorporating D2D and eRRH communication modes. They proposed utilizing MARL and cross-layer network coding techniques to maximize the overall system sum rate. 

\subsection{Motivation and Contributions}

In all previous works, a fronthaul link between the eRRHs and the MBS was assumed. However, in emerging practical applications for dense networks, connecting the eRRHs to the MBS using fronthaul links may not be feasible due to hardware limitations and cost constraints. For example, regarding optical fibre eRRH-MBS links, installing multiple fibres may be financially unbearable for many practical applications or even hard due to harsh geographical terrains. On the other hand, wireless eRRHs-MBS links may cause error propagation losses at far distances since buildings or trees often act as link blockers. 
In view of this, we introduce the concept of using cost-effective elements, referred to as smart helpers (SHs), to extend the traditional F-RANs with no fronthaul connections between the SHs with the MBS. The proposed system is referred to SH-aided F-RAN system. In the SH-aided F-RAN, we consider that a few number of eRRHs are deployed and directly connected to the MBS using fronthaul links, while the SHs are not using fronthaul links to the MBS. Specifically, the SHs can efficiently listen to the communications between the eRRHs and the users in their vicinity without the need for fronthaul links. 
In conventional F-RANs, where only the eRRHs are deployed, the users’ requests can be delivered from the eRRHs in a collaborative way. However, this is not enough in the proposed system since the SHs can cache some popular users’ requested services. Hence, developing caching strategies for both eRRHs and SHs and prioritizing the association of users with them for an effective file delivery becomes quite challenging. To this end, our paper aims to fill this gap by introducing the concept of economical SHs while tackling the joint optimization of caching strategy at eRRHs and SHs and user-eRRH/SH priority association. In more detail, the contribution of the paper is summarized as follows:
\begin{itemize}
\item We present the proposed SH-aided F-RAN system, where   the SHs do not need  fronthaul link connections to the MBS but can listen to the communications between nearby eRRHs and users to smartly cache popular content. Then, the SHs use their cached contents to potentially serve users with frequent on-demand service requests.
\item We develop a priority-based user assignment algorithm to improve the cache hit rate of our SH-aided F-RAN. We then develop a novel MARL algorithm, where popular file segments are treated as virtual agents that make binary caching decisions. The reward function for these virtual agents is designed to minimize the average transmission delay taking into account the limited caching capacity of eRRHs and SHs. We also thoroughly analyze the convergence rate and computational complexity of the proposed algorithm.

\item Numerical results of the proposed framework are presented in various scenarios and compared to several benchmark schemes of conventional F-RANs. The presented numerical results show that SHs can significantly alleviate the load on the fronthaul link and reduce the average transmission delay. Interestingly, our findings also reveal that by tolerating a negligible amount of transmission delay, each eRRH can be replaced with a number of compact SHs.
\end{itemize}

The rest of the paper is organized as follows. In Section~\ref{SectionII}, we introduce the concept of SH-aided F-RAN. Section~\ref{SectionIII} formulates the problem of minimizing average content delivery delay within our proposed SH-aided F-RAN framework. In Section~\ref{SectionIV}, we offer the solution by presenting algorithms for resource allocation and cache optimization. Section~\ref{SectionV} presents simulation results to evaluate the performance of the proposed SH-aided F-RAN and the designed algorithm. Finally, Section~\ref{SectionVI} concludes the paper.
\section{Proposed SH-Aided F-RAN} \label{SectionII}
 \begin{figure}[t!]
	\centering
	\includegraphics[width=0.5\textwidth]{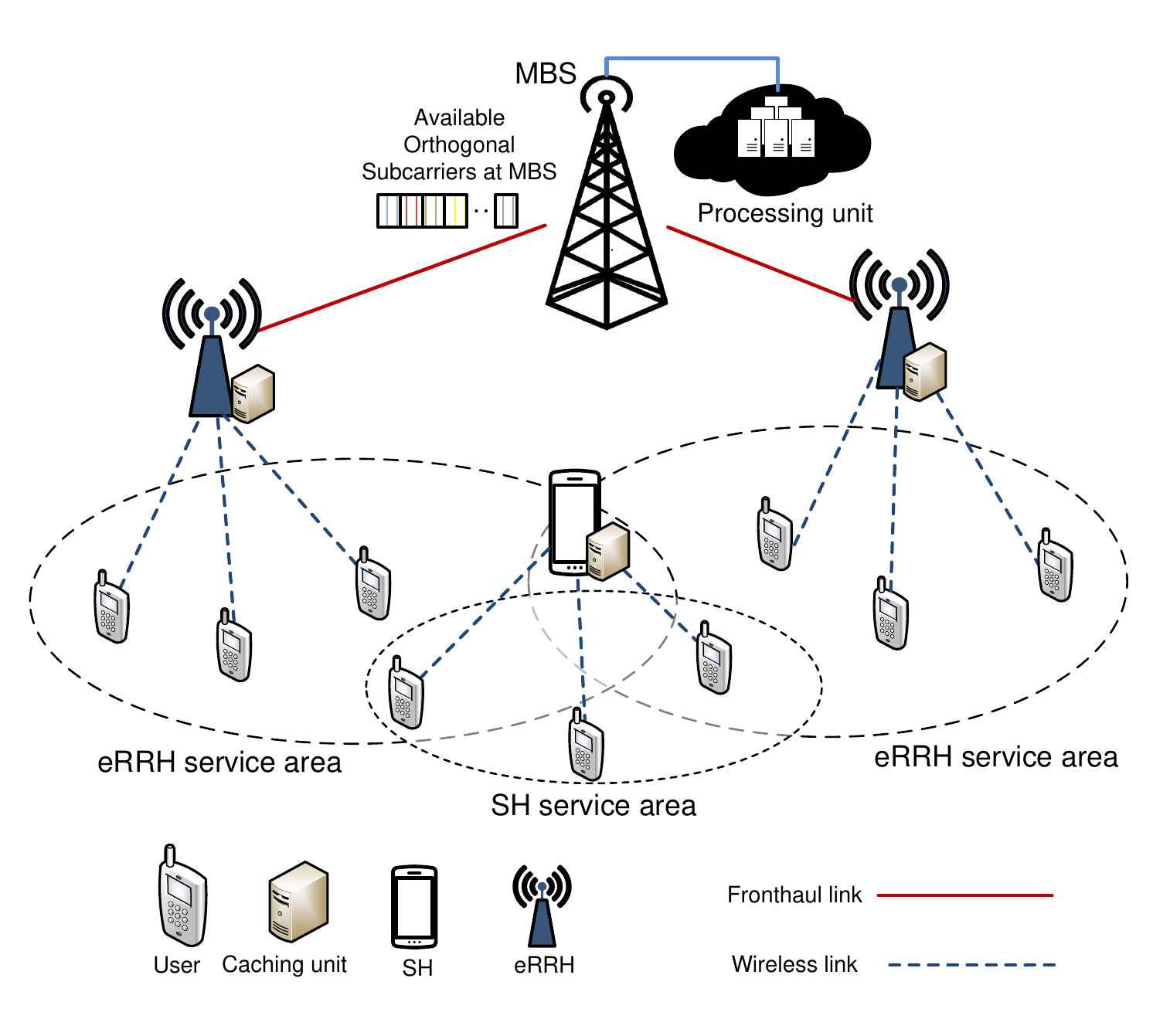}\caption{SH-aided F-RAN architecture.}
	\label{fig1}
\end{figure}

\subsection{Overall Description}
Our proposed SH-aided F-RAN is shown in Fig. \ref{fig1}, where there is a number of $H$ SHs with $\mathcal H= \{1, 2, \cdots, H\}$ denoting the set of all SHs, $S$ eRRHs with $\mathcal S= \{1, 2, \cdots, S\}$ denoting the set of all eRRHs, $U$ users in total with $\mathcal{U} = \{1, \dots, U\}$ denoting the set of all users, and an MBS. We assume that all nodes have a single antenna for simplicity. 
Each SH/eRRH has a certain subset of users under their service area. We represent these subsets by $\mathcal{U}^{\mathcal{H}}_h \subseteq \mathcal{U}$ for the users under the service area of the $h$th SH and $\mathcal{U}^{\mathcal{S}}_s \subseteq \mathcal{U}$ for those under the service area of the $s$th eRRH. 
eRRHs are connected to the MBS using fronthaul links with a limited capacity of $C_{fh}$. They serve users by utilizing their cache or obtaining requested content of users from the MBS as needed.
SHs are relatively cheaper transceivers that have caching capability with a certain amount of caching capacity. We assume that there is no fronthaul link between SHs and the MBS. They listen to the wireless communications between the eRRHs and users happening in their vicinity. Then, they smartly select and cache relevant data and efficiently serve users' requests based on the cached information. 
It is assumed that eRRHs and SHs refine their cache locally to store more on-demand content and serve users effectively.
Furthermore, we assume that SHs are placed randomly and SHs are allowed to decode and cache content. Optimal placement of SHs can lead to improved performance; however, their location optimization is out of the scope of this paper. 
\begin{figure}[t!]
	\centering
	\includegraphics[width=0.5\textwidth]{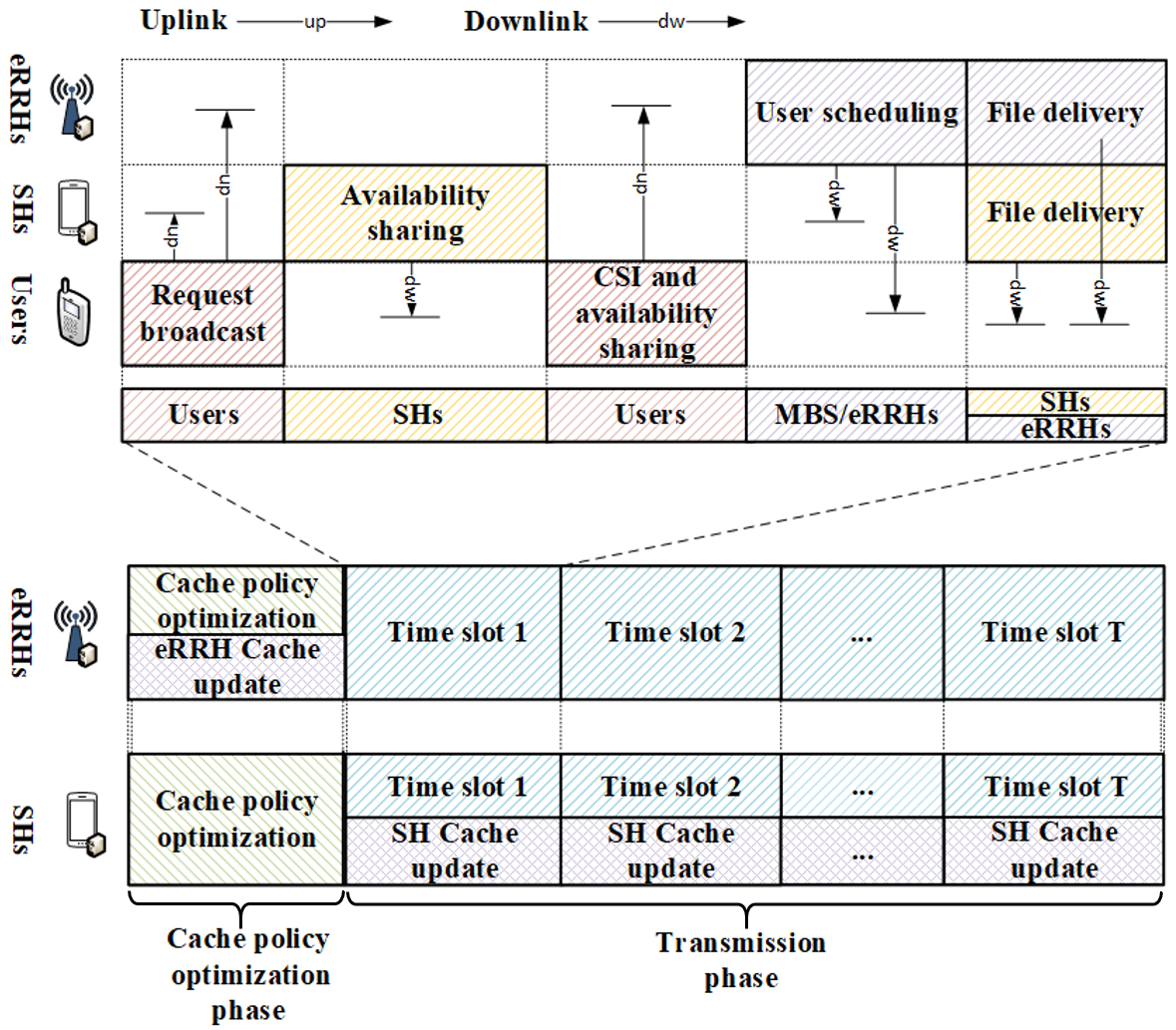}\caption{\textbf{Two-time scale cache optimization and transmission management time frame.}}
	\label{Timeframe}
\end{figure}

We consider a two-time scale cache optimization and transmission management time frame, shown in Fig.~2. At the beginning of each time frame in the cache policy optimization phase, the eRRHs start to optimize their caching policy based on the previous transmission history and then update their cache resources accordingly. Concurrently, SHs optimize their caching policy, with their cache resources being subsequently updated through overhearing the communications in the transmission phase. 
Following the caching policy optimization phase, user requests are addressed during the transmission phase. When users broadcast their requests, SHs inform them about content availability while also users estimate the channel state. Users then share acquired channel state information (CSI) and information regarding requested file availability with eRRHs. Finally, the MBS associates users with various eRRHs/SHs based on their content availability. The MBS communicates user scheduling decisions to eRRHs via the fronthaul links. We consider that the eRRHs and SHs use $R$ orthogonal time/frequency radio resource blocks (RRBs) to serve users. The set of all RRBs is denoted by $\mathcal{R} = \{1, 2, \dots, R\}$. In particular, we consider the F-RAN's resource settings in \cite{9152159, 8736961}.

\begin{figure}[t!]
	\centering
	\includegraphics[width=0.5\textwidth]{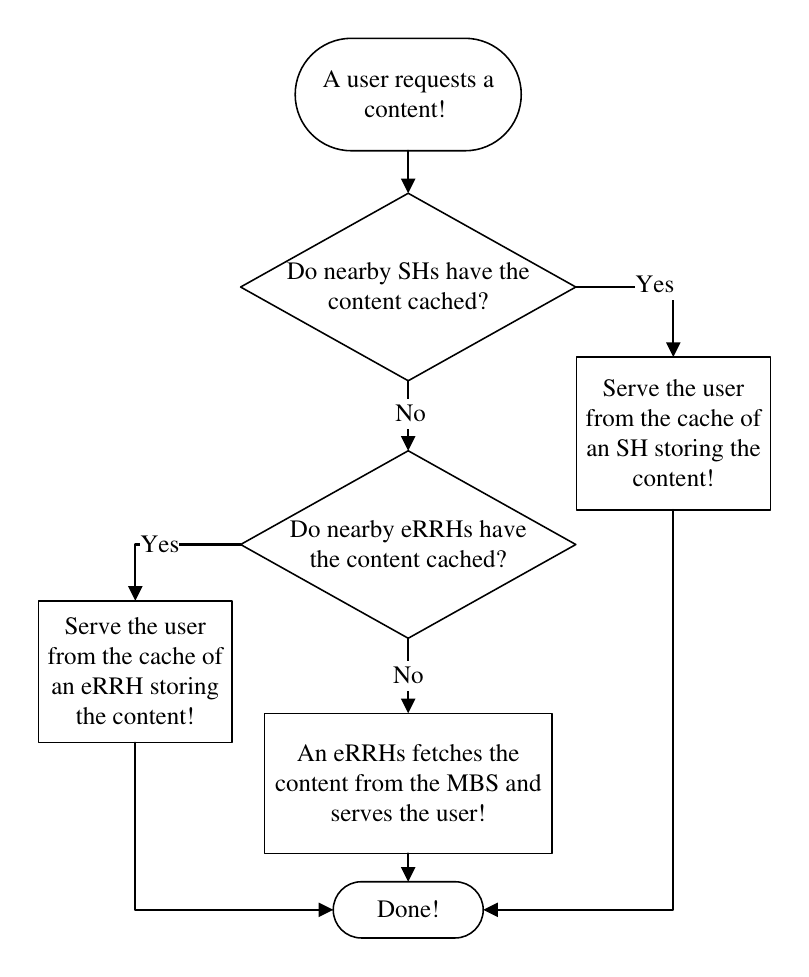}\caption{Content delivery process.}
	\label{flowchart}
\end{figure}

The user scheduling process of SH-aided F-RAN during each time slot is shown in Fig.~\ref{flowchart}. 
Whenever a user requests some content, the MBS associates the user with an SH, caching the requested content. 
In cases where none of the nearby SHs have the requested content in their cache, the system turns to nearby eRRHs to fulfill the user's request. Since the eRRHs are also equipped with caching capabilities, they can directly serve the requested content from their cache to the user without requiring back-and-forth communication with the MBS. However, if none of the nearby SHs or eRRHs have the requested content in their caches, the user will be associated with one of the nearby eRRHs so the eRRH can obtain the content from the MBS and serve the user.
Users are served with different power levels based on the availability of power in SHs and eRRHs, ensuring an efficient content delivery experience.

\subsection{Advantages and Application Scenarios} 
The use of SHs in F-RANs offers numerous advantages for network operators.  SHs are relatively inexpensive compared to eRRHs, and they can be readily deployed without fronthaul links and significant planning effort. As such, it offers a scalable and flexible solution to improve the performance of  F-RANs.  SHs effectively address concerns related to the privacy and power consumption issues associated with CE-D2D communications.  
For example, SHs can be deployed in dense and crowded urban areas. They can reduce network congestion by providing localized caching and enhancing user experience. Also, for public events and other events where people gather for a certain duration, SHs can be deployed easily to serve users from the cached content. In such scenarios, installation of additional eRRHs is time-consuming or may not be feasible. 
Moreover, in rural and remote areas with limited infrastructure, SHs can store frequently accessed content locally, reducing the load on the fronthaul link and improving network performance. In disaster and emergency response situations, SHs can cache critical information and operate as temporary network nodes, ensuring uninterrupted communication. 

\subsection{Operating Constraints}
Let $\mathbf{A}^{\mathcal{S},(t)} $ and $\mathbf{A}^{\mathcal{H},(t)} $ be respectively three-dimensional matrices representing the assignment of RRBs and users to eRRHs and SHs during time slot $t$. The matrix $\mathbf{A}^{\mathcal{S},(t)} $ has dimensions of $U \times R \times S$, where $[\mathbf{A}^{\mathcal{S},(t)}]_{u,r,s} = a^{\mathcal{S},(t)}_{u,r,s}$ indicates if the $u$th user is assigned to the $r$th RRB and associated with the $s$th eRRH. Specifically, $a^{\mathcal{S},(t)}_{u,r,s} = 1$, if the $r$th RRB and the $s$th eRRH are assigned to the $u$th user, and $a^{\mathcal{S},(t)}_{u,r,s}=0$ otherwise. Similarly, the matrix $\mathbf{A}^{\mathcal{H},(t)} $ has a dimension of $U \times R \times H$, where $[\mathbf{A}^{\mathcal{H},(t)}]_{u,r,h} = a^{\mathcal{H},(t)}_{u,r,h}$ represents whether the $u$th user is assigned to the $r$th RRB and the $h$th SH  during time slot $t$, with $a^{\mathcal{H},(t)}_{u,r,h} = 1$ if assigned and $a^{\mathcal{H},(t)}_{u,r,h} = 0$ otherwise.
To avoid interference between users, each RRB is assigned to only one user. Also, due to the radio resource limitation and to ensure fairness among the users, each user is assigned to only one RRB. Hence, the following constraints should be satisfied  
\begin{align}
\sum_{r=1}^R(\sum_{s=1}^S a^{\mathcal{S},(t)}_{u,r,s} + \sum_{h=1}^H a^{\mathcal{H},(t)}_{u,r,h}) \leq 1,    \label{3}\\\
\sum_{u=1}^U(\sum_{s=1}^S a^{\mathcal{S},(t)}_{u,r,s} + \sum_{h=1}^H a^{\mathcal{H},(t)}_{u,r,h}) \leq 1 \label{4}.
\end{align}
Since the number of users is generally higher than the number of available RRBs, we consider serving a certain subset of users during each time slot. Particularly, given a total of $R$ available RRBs in the network, a maximum of $R$ users can be served from the eRRHs and the SHs in a given time slot.

Let $\mathbf{P}^{\mathcal{S},(t)}$ and $\mathbf{P}^{\mathcal{H},(t)}$ be the power allocation matrices of the eRRHs and SHs during time slot $t$. Specifically, $[\mathbf{P}^{\mathcal{S},(t)}]_{s,r} =p^{\mathcal{S},(t)}_{s,r}$ and $[\mathbf{P}^{\mathcal{H},(t)}]_{h,r}=p^{\mathcal{H},(t)}_{h,r}$ are the allocated powers to the $r$th RRB by the $s$th eRRH and the $h$th SH, respectively. Since eRRHs are directly connected to electricity power supplies, their power capabilities are not inherently limited. However, we assume they can transmit with a maximum allowed power on each RRB, denoted by $\bar{P}_\mathcal{S}$, to manage interference between the different cells, leading to improved overall network performance in practice. Nonetheless, in this study, interference from surrounding cells has not been assumed and is considered a potential area for
future research. In addition, SHs are required to properly allocate their available transmit power, limited by $\bar{P}_\mathcal{H}$, to the associated users. Hence, it is essential to consider the following constraints for power allocation
\begin{align}\label{5}
&\sum_{r=1}^R p^{\mathcal{H},(t)}_{h,r} \leq \bar{P}_\mathcal{H} & \forall h=\{1,2,\dots, H\},\\ \label{6}
& p^{\mathcal{S},(t)}_{s,r} =\{  0,\bar{P}_\mathcal{S}\} & \forall s=\{1,2,\dots, S\}.
\end{align}

We consider that the requested content by the users is divided into popular and unpopular file segments. We indicate  $a^{\text{req},(t)}_u$ as the request type parameter of the $u$th user at time slot $t$, where $a^{\text{req},(t)}_u=1$ if the user requests a popular file segment, and $a^{\text{req},(t)}_u=0$ otherwise.  Denote the set of all popular $F$ file segments by $\mathcal F=\{1, 2, \cdots, F\}$, and without loss of generality, assume that each file segment has a size of $B$ bytes. We assume that the popularity of file segments follows well-known Zipf distribution \cite{FRANI8, FRANI11}. According to the Zipf distribution, the popularity of file segment $f$, denoted by $z_f$, is given by 
$z_f = \frac{1}{f^\gamma}/ \sum_{i=1}^{F}\frac{1}{i^\gamma}$,
where $\gamma$ is the Zipf parameter and governs the skewness of popularity distribution, and $\mathcal{Z_{\mathcal{F}}}=\{z_1,z_2,\dots,z_F\}$ represents the popularity distribution of the file segments. We also assume that $\mathcal{F}$ and the popularity of file segments do not change during each time frame, including a large number of time slots. Let $\mathbf{C}^{\mathcal{S}}$ and $\mathbf{C}^{\mathcal{H}}$ be two matrices, respectively representing the caching status of popular file segments in the cache of the eRRHs and SHs during each time frame. The dimension of $\mathbf{C}^{\mathcal{S}}$ is $S \times F$, where $[\mathbf{C}^{\mathcal{S}}] _{s,f} =c^{\mathcal{S},(t)}_{s,f}$ indicates whether the $f$th popular file segment is stored in the cache of the $s$th eRRH at the time slot $t$. Specifically, $c^{\mathcal{S},(t)}_{s,f}=1$ if the segment is cached and $c^{\mathcal{S},(t)}_{s,f}=0$ otherwise. Similarly, the matrix $\mathbf{C}^{\mathcal{H}}$ has a dimension of $H \times F$, where $[\mathbf{C}^{\mathcal{H}}]_{h,f}=c^{\mathcal{H},(t)}_{h,f}$ denotes whether the $f$th file segment is cached in the $h$th SH at the time slot $t$, with $c^{\mathcal{H},(t)}_{h,f} = 1$ if cached and $c^{\mathcal{H},(t)}_{h,f} = 0$ otherwise. We assume  eRRHs and  SHs can cache up to  $K_\text{eRRH}$ and $K_\text{SH}$ file segments, respectively that lead to following constraints
\begin{align}\label{1}
&\sum_{f = 1}^{F} c^{\mathcal{H},(t)}_{h,f} \leq K_\text{SH} & \forall h = \{1,2,\dots,H\},\\\label{2}
&\sum_{f = 1}^{F} c^{\mathcal{S},(t)}_{s,f} \leq K_\text{eRRH} & \forall s = \{1,2,\dots,S\}.
\end{align}
We use the matrices $\mathbf{G}^{\mathcal{S},(t)}$ and $\mathbf{G}^{\mathcal{H},(t)}$ to denote the channel gains at time slot $t$ from users to eRRHs and SHs, respectively. Specifically, we define $[\mathbf{G}^{\mathcal{S},(t)}]_{u,r,s} = g^{\mathcal{S},(t)}_{u,r,s}$ as the channel gain from the $s$th eRRH to the $u$th user over the $r$th RRB, and $[\mathbf{G}^{\mathcal{H},(t)}]_{u,r,h} = g^{\mathcal{H},(t)}_{u,r,h}$ as the channel gain from the $h$th SH to the same user over the same RRB. 
We assume that the channel remains constant while transmitting a single uncoded or coded file segment but changes between successive transmissions. Thus, the achievable rate for serving the $u$th user is determined by 
\begin{align}
R_u = \sum_{r=1}^R W \log (1+\Gamma^{(t)}_{u,r})
\end{align}
where $W$ is the bandwidth of each RRB, and $\Gamma^{(t)}_{u,r}$ is the SINR received by the $u$th user over the $r$th RRB through an additive white Gaussian noise (AWGN) channel during time slot $t$, which is given by
\begin{align}
\Gamma^{(t)}_{u,r} = \frac{\sum_{h=1}^H a^{\mathcal{H},(t)}_{u,r,h} g^{\mathcal{H},(t)}_{u,r,h} p^{\mathcal{H},(t)}_{h,r} +\sum_{s=1}^S a^{\mathcal{S},(t)}_{u,r,s} g^{\mathcal{S},(t)}_{u,r,s} p^{\mathcal{S},(t)}_{s,r}}{N_0}
\end{align}
where $N_0$ is the noise variance.
\subsection{Notations}
Table I summarizes the common notations used throughout the paper.
\begin{table}[t!]\label{Table1}
\small
\renewcommand{\arraystretch}{1.2}
 \caption{Summary of Notations}
 \centering
 \begin{tabularx}{1\columnwidth}{cX}
\hline 
$S$, $H$ & Numbers of eRRHs and SHs\\
$U$, $R$ & Numbers of users and RRBs\\
$c^{\mathcal{S},(t)}_{s,f}$, $c^{\mathcal{H},(t)}_{h,f}$ & Availability of popular file segment $f$ in the $s$th eRRH and the $h$th SH during a time frame\\
$a^{\mathcal{S},(t)}_{u,r,s}$, $a^{\mathcal{H},(t)}_{u,r,h}$ & Assignment of the $r$th RRB and the $u$th user to the $s$th eRRH and the $h$th SH during time slot~$t$\\
$p^{\mathcal{S},(t)}_{s,r}$, $p^{\mathcal{H},(t)}_{h,r}$ & The allocated power of the $s$th eRRH and the $h$th SH to the $r$th RRB during time slot~$t$\\
$g^{\mathcal{S},(t)}_{u,r,s}$, $g^{\mathcal{H},(t)}_{u,r,h}$ & Channel gain from the $u$th user to the $s$th eRRH and the $h$th SH on the $r$th RRB during time slot~$t$\\
$a^{(n)}_{f,s}$, $b^{(n)}_{f,h}$ & Action of agent $f$ in the $s$th eRRH and the $h$th SH at learning iteration $n$\\
$u^{i,(n)}_{f,s}$, $v^{i,(n)}_{f,h}$ & Estimated utility of the action $i$ for the $f$th file segment in the $s$th eRRH and the $h$th SH at learning iteration $n$\\ 
$\pi^{i,(n)}_{f,s}$, $\eta^{(n),i}_{f,h}$ &Preference of the $s$th eRRH and the $h$th SH to take an action $i$ for the $f$th file segment at learning iteration $n$ \\
$K_\text{eRRH}$, $K_\text{SH}$ & Maximum caching capacity of each eRRH and SH\\
$\bar{P}_\mathcal{S}$ & Maximum allowed transmission power of eRRH \\
$\bar{P}_\mathcal{H}$ & Available transmission power of each SH\\
\hline 
 \end{tabularx}
 \end{table}
\section{Delay Minimization in SH-aided F-RAN}  \label{SectionIII}
The transmission delay of serving the $u$th user with 
a popular file segment of a size $B$ bytes from an SH, denoted by $D^\mathcal{H}_u$, or an eRRH, denoted by $D^\mathcal{S}_u$, is given by $B/R_{u}$. 
When the requested content is unavailable in the SHs or eRRHs, the transmission delay comprises both the time it takes to fetch the content from the MBS and the time to transmit it to the user through the connected eRRH. Hence, the transmission delay for serving the user from the MBS is determined by $D_u^{MBS}=\frac{B}{R_{u}}+\frac{B}{C_\text{fh}}$.  
Accordingly, the average latency for the $u$th user to download its requested content can be expressed as
\begin{align} \label{Du}
D_u= & a^{\text{req},(t)}_u[\sum_{s=1}^S\sum_{r=1}^R a^{\mathcal{S},(t)}_{u,r,s}c^{\mathcal{S},(t)}_{s,f}D^\mathcal{S}_u +\sum_{h=1}^H \sum_{r=1}^R a^{\mathcal{H},(t)}_{u,r,h} c^{\mathcal{H},(t)}_{h,f}D^\mathcal{H}_u 
\nonumber\\&+ \sum_{s=1}^S\sum_{r=1}^Ra^{\mathcal{S},(t)}_{u,r,s}(1-c^{\mathcal{S},(t)}_{s,f})D^{MBS}_u] 
\nonumber\\&+(1-a^{\text{req},(t)}_u) D^{MBS}_u.
\end{align}
Since the maximum number of served users is equal to the number of the available RRBs, the average file segments downloading delay of all the users can be obtained by $D_\text{ave}~=~\frac{1}{R}~\sum_{u=1}^U D_u$.
Now, we formulate an optimization problem that minimizes average content delivery delay by considering user scheduling and cache optimization subject to the constraints~(\ref{3})--(\ref{2}) as follows.
\begin{align} \label{Problem}
	&\min_{\underset{ \mathbf{C}^{\mathcal{H}},\mathbf{A}^{\mathcal{H},(t)},\mathbf{P}^{\mathcal{H},(t)}}{\mathbf{C}^{\mathcal{S}},\mathbf{A}^{\mathcal{S},(t)},\mathbf{P}^{\mathcal{S},(t)}}} \frac{1}{R} \sum_{u=1}^U D_u\nonumber\\
	&~~~~~~~\textrm{s.t.}   \begin{array}{clc}
    &\textrm{C1: }\sum_{r=1}^R(\sum_{s=1}^S a^{\mathcal{S},(t)}_{u,r,s} + \sum_{h=1}^H a^{\mathcal{H},(t)}_{u,r,h}) \leq 1,~~\\\vspace{0.1cm}
    &\textrm{C2: }\sum_{u=1}^U(\sum_{s=1}^S a^{\mathcal{S},(t)}_{u,r,s} + \sum_{h=1}^H a^{\mathcal{H},(t)}_{u,r,h}) \leq 1,~~\\\vspace{0.1cm}
    &\textrm{C3: } \sum_{f=1}^F c^{\mathcal{H},(t)}_{h,f} \leq K_\text{SH},\\\vspace{0.1cm}
    &\textrm{C4: }\sum_{f=1}^F c^{\mathcal{S},(t)}_{s,f} \leq K_\text{eRRH},\\\vspace{0.1cm}
    &\textrm{C5: } a^{\mathcal{S},(t)}_{u,r,s}, a^{\mathcal{H},(t)}_{u,r,h} \in \{0,1\},\\
    &\textrm{C6: }\sum_{r=1}^R p^{\mathcal{H},(t)}_{h,r} \leq \bar{P}_\mathcal{H},\\\vspace{0.1cm}
    &\textrm{C7: } p^{\mathcal{S},(t)}_{s,r}  =\{0,\bar{P}_\mathcal{S}\},\\
    &\textrm{C8: } c^{\mathcal{S},(t)}_{s,f},c^{\mathcal{H},(t)}_{h,f}, \in \{0,1\},
    \end{array} 
\end{align}
where constraints C1 and C2 ensure a one-to-one mapping between the users and the RRBs such that each user is assigned to only one RRB, and each RRB is assigned to only one user. Constraints C3 and C4 limit the caching capacity of the SHs and the eRRHs to $\Gamma_{SH}$ and $\Gamma_{eRRH}$ file segments, respectively.
Constraint C6 specifies a maximum available transmission power for the SHs, while constraint C7 sets a limit on the maximum transmission power on each RRB in the eRRHs. Constraint C5 and C8 consider binary decisions for variables $a^{\mathcal{S},(t)}_{u,r,s}$, $a^{\mathcal{H},(t)}_{u,r,h}$, $c^{\mathcal{S},(t)}_{s,f}$, and $c^{\mathcal{H},(t)}_{h,f}$.
The optimization variables  are the cache matrices for both SHs and eRRHs, assignment vectors that indicate which user is assigned to which RRB in each eRRH and SH, and the power allocation vectors for both eRRHs and SHs.

The derived optimization problem in  (\ref{Problem}) is a mixed-integer nonlinear program (MINLP). It is generally difficult to obtain the globally optimal solution of a MINLP.  
In fact, due to its inherent computational complexity resulting from the interdependence of variables $c^{\mathcal{S},(t)}_{s,f}$, $c^{\mathcal{H},(t)}_{h,f}$, $a^{\mathcal{S},(t)}_{u,r,s}$, and $a^{\mathcal{H},(t)}_{u,r,h}$,  solving this problem is computationally intractable.  To tackle this intractability, we propose to solve (\ref{Problem}) iteratively. Specifically, the optimization problem is first divided into two subproblems:  1) user assignment and power allocation and 2) cache resource optimization. Then, each subproblem is separately optimized. In the next section, each subproblem and the corresponding solution will be discussed in detail. We also analyze their convergence and computational complexity.

\section{Proposed solution: Resource Allocation and Cache Optimization}\label{SectionIV}
\subsection{User Assignment and Power Allocation} \label{usera}
In this subsection, we effectively schedule users to eRRHs and SHs to minimize the average transmission delay through the following subproblem 
\begin{align} 	\min_{\mathbf{A}^{\mathcal{H},(t)},\mathbf{A}^{\mathcal{S},(t)},\mathbf{P}^{\mathcal{H},(t)},\mathbf{P}^{\mathcal{S},(t)}} ~~~\sum_{u=1}^U D_u, 
	~~~\textrm{s.t.~C1,~C2,~C5--C8}.
\end{align}
We construct a conflict graph that includes all the possible user assignment states to consider constraints C1 and C2. Then, we present an algorithm to associate the users with SHs and eRRHs by finding the graph's minimum weight independent set (MWIS). Additionally, this algorithm optimizes the allocation of transmission power between users. The obtained user association and power allocation solution minimizes the average transmission delay of the network. 

Consider $\mathcal{G} = (\mathcal{V}^{\mathcal{H}},\mathcal{V}^{\mathcal{S}},\mathcal{E}, \mathcal{W})$ as a weighted undirected conflict graph, for which $\mathcal{V}^{\mathcal{H}}$ and $\mathcal{V}^{\mathcal{S}}$ are the sets of vertices of the SHs and eRRHs, respectively. The graph has edges represented by $\mathcal{E}$, and weights assigned to the vertices are given by $\mathcal{W}$. Let $v^{\mathcal{H}} = (h, u, r)$ represent the vertices belonging to $\mathcal{V}^{\mathcal{H}}$, where each vertex comprises an SH $h \in \mathcal{H}$, a user $u \in \mathcal{U}^{\mathcal{H}}_h$, and an RRB $r \in \mathcal{R}$. Similarly, $v^{\mathcal{S}} = (u, r, s)$ represents vertices of $\mathcal{V}^{\mathcal{S}}$, where each vertex includes an eRRH $s \in \mathcal{S}$, a user $u \in \mathcal{U}^{\mathcal{S}}_s$, and an RRB $r \in \mathcal{R}$.
Every two vertices have a connecting conflict edge if they include the same user or the same RRB. Hence, the vertices of $\mathcal{V}^{\mathcal{H}}$ and $\mathcal{V}^{\mathcal{S}}$ are also linked to each other if they contain the same user or RRB. 
For instance, consider two sample vertices: $v^{\mathcal{H}} = (u, r, h)$ and $v^{\mathcal{S}} = (u', r', s)$. These two vertices are connected to each other if either $r = r'$, $u = u'$, or both conditions are met.
The weight of each vertex is the transmission delay of serving the associated user of that vertex by the corresponding RRB and SH/eRRH.
We prioritize SHs for serving users. For users requesting a specific file segment not present in SHs, we designate eRRHs to serve to them. In this way, we reduce the average transmission delay by maximizing the utilization of SHs' cached resources to serve a larger number of users.

\begin{algorithm}[t!]
 \caption{User Assignment and Power Allocation Algorithm}
 \begin{algorithmic}[1]
 \setstretch{1.1}
 \renewcommand{\algorithmicrequire}{\textbf{input:}}
 \renewcommand{\algorithmicensure}{\textbf{output:}}
 \label{Algorithm1}
 \REQUIRE $\mathbf{C}^{\mathcal{S}},\mathbf{C}^{\mathcal{H}},\mathbf{G}^{\mathcal{S},(t)},\mathbf{G}^{\mathcal{H},(t)}$
 \ENSURE  $\mathbf{A}^{\mathcal{H},(t)},\mathbf{A}^{\mathcal{S},(t)},\mathbf{P}^{\mathcal{H},(t)},\mathbf{P}^{\mathcal{S},(t)}$
 \STATE \textbf{Graph Formation:}
  \FOR {$r = 1:R$}
  \FOR {$h = 1:H$}
  \FOR {$u = 1:\mathcal{U}^{\mathcal{H}}_h$}
  \STATE initiate a vertex $v^{\mathcal{H}}= (u,r,h)$
  \STATE $\mathbf{w}(u,h,r) = D_u$ using (\ref{Du})
  \ENDFOR
  \ENDFOR
  \FOR {$s = 1:S$}
  \FOR {$u = 1:\mathcal{U}^{\mathcal{S}}_s$}
  \STATE initiate a vertex $v^{\mathcal{S}}= (u,r,s)$
  \STATE $\mathbf{w}(u,s,r) = D_u$  using (\ref{Du})
  \ENDFOR
  \ENDFOR  
  \ENDFOR
  \STATE Establish connecting edges between vertices sharing the same RRB and/or user.
  \STATE \textbf{User Association:}
  \STATE MWIS($\mathcal{V}^{\mathcal{H}}$)
  \STATE MWIS($\mathcal{V}^{\mathcal{S}}$)
  \STATE \textbf{Power Allocation:}
  \FOR {$s = 1:S$}
  \FOR {$r = 1:R$}
  \STATE $p^{\mathcal{S},(t)}_{s,r}=\bar{P}_\mathcal{S}\sum_{u=1}^U a^{\mathcal{S},(t)}_{u,r,s}$
  \ENDFOR
  \ENDFOR
  \FOR {$h = 1:H$}
  \STATE Power allocation of the $h$th SH using \textbf{WF} algorithm
  \ENDFOR
 \end{algorithmic} 
 \end{algorithm} 
We propose the following procedure for user scheduling. 
First, we construct the graph by generating all the possible vertices of the eRRHs and the SHs, calculating their corresponding weights, and making the above-mentioned conflict links.
Second, we need to find the MWIS of the whole graph such that we assign the users first to the SHs, and the remaining users need to be assigned to the eRRHs. Since this process is intractable, we use a heuristics method that was used in \cite{9152159} to find a suboptimal user scheduling solution. In this fashion, first, we find the vertex of the SHs with the minimum weight, assign the user of that vertex to its RRB and SH, and remove all of the vertices linked to that vertex. This process is repeated till there are no more SH vertices in the graph. Then, the same procedure will be applied to the remaining vertices of  $\mathcal{V}^{\mathcal{S}}$, so that all of the RRBs are assigned to the users. 
Third, to allocate the power resources among the eRRHs effectively, since each RRB is assigned to only one user and the RRBs are orthogonal, we grant $\bar{P}_\mathcal{S}$ to the RRBs used by an eRRH. However, due to the limited amount of power in each SH during a time slot, a Water-Filling (WF) algorithm is utilized at each SH to allocate the power between users properly. Such a power allocation procedure takes into account the number of users served by each SH and their respective power requirements to ensure that each user receives an appropriate amount of power. This ensures that the power is utilized effectively and that each user is able to receive its requested file segment with a possible minimum delay.
To summarize, the MBS takes the following three steps to schedule the users: (i) graph construction, (ii) user association, and (iii) power allocation, which are presented in Algorithm~\ref{Algorithm1}.
\subsection{Cache Optimization} \label{cacheS}
We consider the following subproblem to optimize the cache state of the eRRHs and SHs
\begin{align} \label{Problem1}
\min_{\mathbf{C}^{\mathcal{S}},\mathbf{C}^{\mathcal{H}}} ~~~\sum_{u=1}^U D_u, 
	~~~\textrm{s.t.~C3,~C4,~C8}.
\end{align}
For each eRRH and SH, there exist permutations of $K_\text{eRRH}$ and $K_\text{SH}$, respectively, out of a total of $F$ possible states. This results in a number of potential solutions for the problem, specifically $[P(F, \Gamma_{eRRH})]^S \times [P(F, \Gamma_{SH})]^H$, where $P(n,r) = n! / (n-r)!$ represents the permutation of selecting $r$ options out of $n$ possible states. Given the considerable computational complexity involved, we develop two separate RL algorithms to update the cache resources of eRRHs and the SHs.

\subsubsection{Cache Updates in the eRRHs} 
eRRHs update their cache resources by fetching their desired file segments from the MBS. The more file segments saved in each eRRH, the more the chance of serving users from the cache of their assigned eRRH. However, the cache capacity of each eRRH is limited by $K_\text{eRRH}$. In this regard, we develop an actor-critic MARL algorithm to optimize the cache resources of the eRRHs. 
Using the actor-critic algorithm, agents learn both a policy function and a value function referred to as the actor and critic, respectively. During each iteration, the critic evaluates and updates the actor's policy function by predicting the future rewards. Following this, the actor takes an action based on its updated policy function and receives a reward. The critic then updates its value function based on the received reward and the prediction error \cite{RL_book}.
In our proposed algorithm, $F$ virtual agents are created in each eRRH, with each agent corresponding to one of the popular file segments. These virtual agents make binary decisions based on each popular file segment's learned caching policies. 

Since maximizing the cache hit rate at the eRRHs leads to minimizing the average transmission delay of serving the users, the rewards of the virtual agents are considered to be related to their cache hit rate. Besides, the virtual agents should avoid caching the file segments being requested less because of the limited caching capacity. Accordingly, we consider the reward of virtual agent $f$ of eRRH $s$ as follows 
\begin{align}
    \mu^{(n)}_{f,s} = \alpha_{\mu} [(1-2 c^{\mathcal{S},(n)}_{s,f}) + c_l~l(f,s)]
\end{align}
where $(1-2 c^{\mathcal{S},(n)}_{s,f})$ is the cost of caching the file segment $f$, $c_l$ is the impact of the cache hit status, and $l(f,s)$ is the legitimacy of caching the file segment $f$ and considered as follows. If the file segment $f$ is requested by a user assigned to the $s$th eRRH, $l(f,s) = -1$ in case of cache hit failure and $l(f,s) = +1$ if a cache hit success happens, otherwise $l(f,s) = 0$.
Denote $b^{\mathcal{S},(n)}_{f,s}$ as the action of the $f$th virtual agent in the $s$th eRRH at iteration $n$, and parameterize its policy function as an exponential softmax distribution
\begin{align}
 \beta^{i,(n)}_{f,s}(u^{i,(n)}_{f,s}) = \frac{\exp(\lambda_p u^{i,(n)}_{f,s})}{\exp(\lambda_p u^{0,(n)}_{f,s})+\exp(\lambda_p u^{1,(n)}_{f,s})}
\end{align}
where $u^{i,(n)}_{f,s}$ is the estimated utility of action $i$ at iteration $n$ and $\lambda_p$ is the exploration-exploitation factor. 
During each iteration step, every eRRH takes a random action generated from $\pi^{i,(n)}_{f,s}$ to cache $K_\text{eRRH}$ files and obtains its own reward $\mu^{(n)}_{f,s}$. Obtaining the reward, the policy and the utility functions are updated as follows
\begin{align} \label{UP1}
&u^{i,(n+1)}_{f,s} = u^{i,(n)}_{f,s} + \alpha_u  \left( \mu^{(n)}_{f,s} - u^{i,(n)}_{f,s} \right)\mathcal{I}\{b^{\mathcal{S},(n)}_{f,s} =i\}\\
&\pi^{i,(n+1)}_{f,s} = \pi^{i,(n)}_{f,s} + \alpha_{\pi} \left [\beta^{i,(n)}_{f,s}(u^{i,(n)}_{f,s}) - \pi^{i,(n)}_{f,s}\right ].
\end{align}
The learning rates $\alpha_u$ and $\alpha_{\pi}$ represent the rates at which the utility and policy functions are updated, respectively. To ensure the convergence of the algorithm, these learning rates must satisfy specific conditions \cite{Game}:
\begin{align}
\label{Convergence1}
    \sum_{n= 1}^{+\infty} \alpha_u = +\infty, \sum_{n= 1}^{+\infty} \alpha_u^2 \leq +\infty,\\\label{Convergence2}
    \sum_{n= 1}^{+\infty} \alpha_{\pi} = +\infty, \sum_{n= 1}^{+\infty} \alpha_{\pi}^2 \leq +\infty,\\\label{Convergence3}
    \lim_{n\rightarrow + \infty} \frac{\alpha_{\pi}}{\alpha_{u}} = 0.~~~~~~~~~~~~~~~~~~~
\end{align}
\subsubsection{Cache Updates in the SHs}
Similarly, $F$ virtual agents are created in each SH, where each agent is dedicated to a specific popular file segment. In contrast to the virtual agents of eRRHs, these agents make binary caching decisions by considering the file segments received during the previous time slot or those already stored in the SH's cache. However, the critic operates the same as the virtual agents in eRRHs. Denote the action of the $f$th virtual agent in the $h$th SH and its reward signal at iteration $t$ as $b^{\mathcal{H},(n)}_{f,s}$ and $\nu^{(n)}_{f,h}$, respectively. The reward signal of SH virtual agents is considered as follows
\begin{align}
    \nu^{(n)}_{f,h} = \alpha_{\nu} [(1-2 c^{\mathcal{H},(n)}_{h,f}) + c_l~l(h,s)].
\end{align}
Then $ v^{i,(n)}_{f,h}$ and $\eta^{i,(n)}_{f,s}$ as the utility and policy functions of the virtual agent are respectively updated based on the decision of the actor as follows
\begin{align}\label{UP2}
&v^{i,(n+1)}_{f,h} = v^{i,(n)}_{f,h} + \alpha_v  \left( \nu^{(n)}_{f,s} - v^{i,(n)}_{f,s} \right)\mathcal{I}\{b^{\mathcal{H},(n)}_{f,s} =i\},\\
&\eta^{i,(n+1)}_{f,s} = \eta^{i,(n)}_{f,h} + \alpha_{\eta} \left [\beta^{i,(n)}_{f,h}(v^{i,(n)}_{f,s}) - \eta^{i,(n)}_{f,s}\right ], 
\end{align}
where $\alpha_v$ and $\alpha_{\eta}$ are the learning rates of the utility function and the policy which should satisfy the conditions (\ref{Convergence1})--(\ref{Convergence3}) for the convergence of the algorithm.
\begin{algorithm}[t!]
 \caption{User Assignment and Caching Algorithm}
        \begin{algorithmic}[1]
\setstretch{1.1}
 \renewcommand{\algorithmicrequire}{\textbf{Input:}}
 \renewcommand{\algorithmicensure}{\textbf{Output:}}
 \label{Algorithm2}
  \STATE \textit{\textbf{1. Cache Policy Learning Stage}}
  \STATE Set $u^{i,(n)}_{f,s}=v^{i,(n)}_{f,h}= 0$ and $\pi^{i,(n)}_{f,s} = \eta^{(n),i}_{f,h} = 0.5$ for all agent
  \FOR{ $n = 1 : N$ }
  \FOR{ $f = 1 : F$ }
  \FOR{ $s = 1 : S$ }
  \STATE Generate an action $b^{\mathcal{S},(n)}_{f,s} \sim \pi^{i,(n)}_{f,s}$
  \ENDFOR
  \FOR{ $h = 1 : H$ }
  \STATE Generate an action $b^{\mathcal{H},(n)}_{f,h} \sim \eta^{i,(n)}_{f,h}$
  \ENDFOR
  \ENDFOR
  \STATE Obtain the reward of each agent, $\mu^{(n)}_{f,h}$ and $\nu^{(n)}_{f,h}$
  \STATE Update $u^{i,(n+1)}_{f,s}$ and $\pi^{i,(n+1)}_{f,s}$ using (\ref{UP1})
  \STATE Update $v^{i,(n+1)}_{f,h}$ and $\eta^{i,(n+1)}_{f,h}$ using (\ref{UP2})
  \STATE Update the cache of the eRRHs and the SHs by $b^{\mathcal{S},(n)}_{f,s}$ and $b^{\mathcal{H},(n)}_{f,h}$
  \ENDFOR
  \STATE \textit{\textbf{2. Cache Resource Update Stage}}
  \STATE Update the cache of the eRRHs and the SHs by $b^{\mathcal{S},(N)}_{f,s}$ and $b^{\mathcal{H},(N)}_{f,h}$
  \STATE \textit{\textbf{3. User Assignment and Power Allocation Stage }}
  \FOR{ $t = 1 : T$ }
  \STATE Assign the users to the eRRHs and the SHs and allocate the power to them using Algorithm \ref{Algorithm1}
 \ENDFOR
 \end{algorithmic} 
 \end{algorithm}
 \subsection{Summary Of The Overall Solution}
 The algorithm proposed to minimize the average transmission delay can be outlined as follows. At each iteration of the algorithm, the initial step involves updating the cache resources of the eRRHs, which generates a decision $b^{\mathcal{S},(n)}_{f,s}$. This decision is created based on the policy $\pi^{i,(n)}_{f,s}$, and it aims to allocate popular file segments to the eRRHs' cache resources. The SHs use their policy, which takes into account the latest $\eta^{i,(n)}_{f,s}$ and previous received communications, to generate a decision $b^{\mathcal{H},(n)}_{f,h}$. The decision is based on whether the file segments have already been cached or were just received in the previous step. 
 After updating the cache of both eRRHs and SHs, we employ Algorithm \ref{Algorithm1} to allocate the RRBs and users to the respective eRRHs and SHs. The eRRHs grant their assigned RRBs to $\bar{P}_\mathcal{S}$, while the SHs use the WF algorithm to distribute the available power among the assigned RRBs. Each agent receives a reward based on its cache hit factor, and their policies are updated using equations (\ref{UP1}) and (\ref{UP2}) for the next iteration step.
 The learning process continues until it reaches the maximum number of iterations, $n =N$. After $N$ learning iterations, eRRHs update their cache resources based on the most updated policy by fetching the content from the MBS. However, the SHs update their cache resources during the transmission phase while it is overhearing the communications around them.
 The process of union user scheduling and caching of both the SHs and eRRHs is illustrated in Algorithm \ref{Algorithm2}.
\subsection{Convergence and Complexity of the Algorithm}
\subsubsection{Convergence}
The process described in Algorithm \ref{Algorithm2} updates the probability of the agents of eRRHs and SHs selecting actions at each iteration. When the algorithm converges, the probabilities become stable and do not change much. This can be shown by the following theorem.
 \subsubsection*{Theorm 1}
 \textit{The policies of binary action selection for each agent, i.e. $\pi^{i,(n)}_{f,s}$ and $\eta^{i,(n)}_{f,s}$ is converged as $\lim_{n \to  \infty} \pi^{i,(n)}_{f,s} \to  \bar{\pi}^{i}_{f,s} $ and $\lim_{n \to \infty} \eta^{(n),i}_{f,h} \to \bar{\eta}^{i}_{f,h} $, $\forall i \in \{1,2\}, f \in \{1,2,\dots,F\}$, $s \in  \{1,2,\dots,S\}$, and $h \in  \{1,2,\dots,H\}$. Here
 \begin{align}
    \bar{\pi}^{i}_{f,s} = \frac{\exp(\lambda_p \bar{\mu}_{f,s}^i)}{\sum_{i=1}^2 \exp(\lambda_p \bar{\mu}_{f,s}^i)}\\
    \bar{\eta}^{i}_{f,h} = \frac{\exp(\lambda_p \bar{\nu}_{f,h}^i)}{\sum_{i=1}^2 \exp(\lambda_p \bar{\nu}_{f,h}^i)}
 \end{align}
 where $\bar{\mu}_{f,s}^i$ and $\bar{\nu}_{f,h}^i$ are expected reward of taking action $i$ by the $f$th agent of the $s$th eRRH and the $h$th SH, respectively.}
 
 For proof, please refer to the Appendix. 
 \subsubsection{Complexity of Algorithm}
In algorithm 1, for user assignment during each transmission time slot, a graph with all cases of eRRHs and SHs vertices should be constructed, which has a computational complexity of $ \mathcal{O}(R(S+H)U)$. To generate the conflict edges between the vertices of the graph, we are required to check each vertex to the other vertices to see if they share the same RRB, the same user, or both. Hence, $\mathcal{O}(R^2U^2(S+H)^2)$ operations are needed. In addition, the vertex weight calculation and traversal are performed within the specified timeframe by the users and transferred to the eRRHs and the MBS, which mitigates potential complexity and delays caused in Algorithm~1. To find the minimum weighted independent set of the constructed graph, the vertex with the minimum weight out of $R(S+H)U$ vertices needs to be found for $R$ times. So, the computation complexity of this stage is generally bounded by $\mathcal{O}(UR^2(S+H))$.  Hence, the total complexity of user assignment is $\mathcal{O}(R(S+H)U+R^2U^2(S+H)^2+UR^2(S+H)) = \mathcal{O}(R^2U^2(S+H)^2)$ operations.
In the power allocation stage, the SHs allocate the available power to the assigned users and RRBs based on WF algorithm. Where the computational complexity of WF for $N$ channels is generally in the range of $\mathcal{O}(N \log N)$ [41]. The number of users that each SH can serve is bounded to a small number due to the limited available transmission power. So, the computational complexity order of this stage is approximately $\mathcal{O}(H)$.

For the learning stage, the cache resource update stage requires generating $K_\text{eRRH}$ and $K_\text{SH}$ random values from $\pi^{(n),i}_{f,s}$ and $\eta^{(n),i}_{f,h}$. Using the Alias method [40], the computational complexity of the first stage at most is $\mathcal{O}(F (S+H))$. As the learning update stage involves a fixed number of operations, i.e., $2F(S+H)$ exponents, multiplications and additions, the computational complexity of the last stage in each learning step is $\mathcal{O}(3F(S+H))$. Therefore, the overall worst-case computational complexity of the learning stage for the entire $N$ learning iterations is approximately $\mathcal{O}(3 N F(S+H))$.
 
\section{Numerical Results and Analysis} \label{SectionV}
We numerically evaluate the performance of our developed scheme of the envisioned SH-aided F-RAN in terms of average transmission delay, average load on the fronthaul link, and cache hit rate with benchmark schemes proposed for conventional F-RANs. In particular, we compare the performance of the proposed SH-aided F-RAN with a similar network with the same number of CE-relays instead of SHs. Moreover, we quantify the performance gain that can be achieved using the developed caching algorithm considering four different benchmark schemes: 1) random caching strategy, 2) most probable caching scheme (MPC), 3) two hybrid caching schemes of MPC and random, and 4) a uniform caching scheme with considering a non-overlapping most probable caching strategy for the SHs where each SH caches unique file segments which are not cached at the other SHs.
\begin{figure}
        \centering
        \includegraphics[width=\columnwidth]{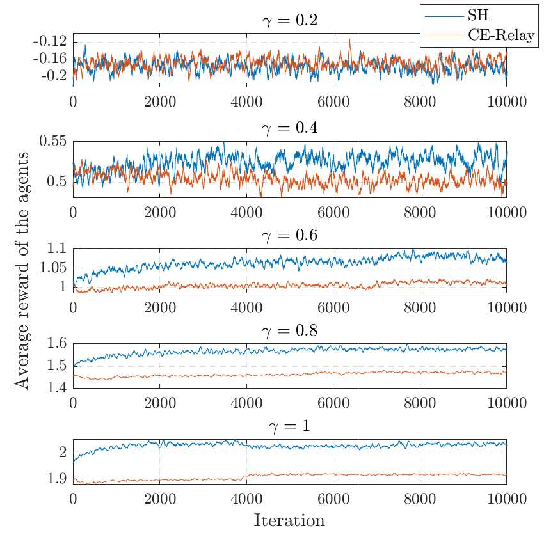}
        \caption{Average reward of agents of one SH and one CE-Relay at each training iteration for different values of the Zipf's parameter $\gamma$.}
        \label{Fig2}
\end{figure}
\subsection{Basic Simulation Settings}
We assume the eRRHs, SHs, and users are uniformly located within a circle of 1.5~km radius. The minimum distance between neighbouring eRRHs is set to 300~m and the minimum distance between neighbouring SHs is set to 150~m. The MBS is assumed to have 50 orthogonal RRBs over each time frame to serve the users. It is assumed that there are 10,000 popular file segments, each with the size of $1.25$~MBytes, where their popularity follows the Zipf distribution with different values of the Zipf parameter. Users are assumed to request a popular file segment with a probability of 0.3. Each of the eRRHs and SHs can store/cache up to 200 and 100 file segments, respectively. 
The channel gains between users and eRRHs, $g^{\mathcal{S},(t)}_{u,r,s}$, the channel gains between users and SHs, $g^{\mathcal{H},(t)}_{u,r,h}$ are all assumed to have a fading model used in the literature, e.g., \cite{8736961, FRANI8}, that consists of  1) path-loss of $128.1+37.6 \log10(dis.[km])$, 2) log-normal shadowing with $4dB$ standard deviation, and 3) Rayleigh fading with zero-mean and unit variance.   
The additive white Gaussian (AWGN) noise power and the fronthaul link capacity are assumed to be $N_0 = -174~dBm$ and $C_{fh} = 12.5$ MBytes/sec, respectively. The maximum allowed transmission power of each eRRH over an RRB is $\bar{P}_S = 7$ mW. In addition, the maximum available power at each SH is assumed to be $\bar{P}^{max}_H = 200$ mW that is allocated to different RRBs using the WF algorithm. The impact of the cache hit rate on the reward is considered to be $c_l = 2$. We evaluate each metric by averaging over 2000  Monte-Carlo simulation runs.

\subsection{Evaluation and Comparison}


Figure~\ref{Fig2} presents the average reward of the virtual agents of one SH and a CE-relay in the network over 10,000 iterations in different values of Zipf's parameter $\gamma$.  
As seen, the rewards of SH-aided F-RANs converge faster to the maximum value compared to CE-relay-aided F-RANs. This is because of the capability of SHs to overhear ongoing communications in their vicinity, thereby gaining access to more diverse communication information. In contrast, CE-relays acquire information solely when engaged in content delivery from the eRRHs to the users. Based on the results, we choose 2,000 iterations as the stopping point for our learning algorithm to balance learning efficiency and policy stability. By doing so, we ensure that the algorithm achieves an optimized caching policy without unnecessary computational overhead and extended training time.


\begin{figure}
        \centering
        \includegraphics[width=\columnwidth]{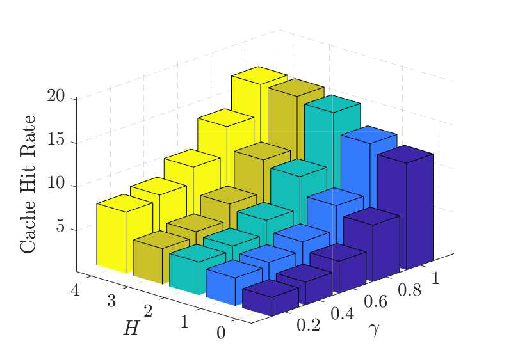}
        \caption{Total cache hit rate using the proposed algorithm for $S = 4$ with various number of SHs.}
\label{Fig5}
\end{figure}

\begin{figure}
    \begin{subfigure}[b]{0.5\textwidth}
    \centering
    \includegraphics[width=\columnwidth]{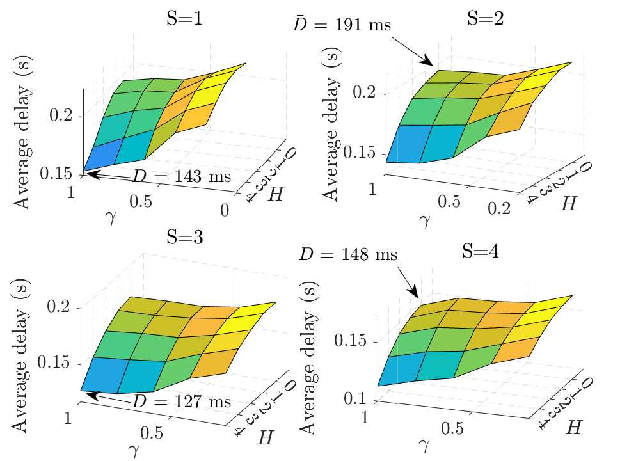}
    \caption{Average transmission delay utilizing different numbers of SHs.}
     \end{subfigure}
    \begin{subfigure}[b]{0.5\textwidth}
    \centering
    \includegraphics[width=\columnwidth]{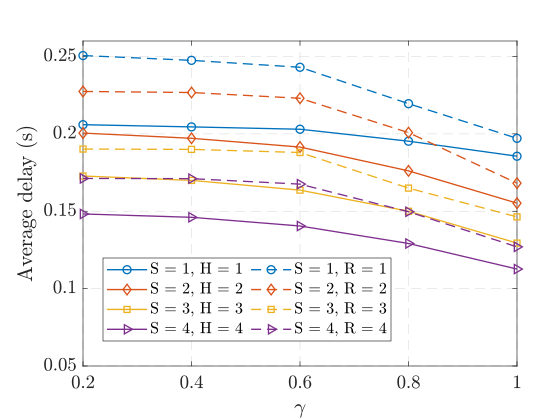}
    \caption{Comparing the average transmission delay of SHs versus the CE-relays.}
     \end{subfigure}
    \caption{Average transmission delay of different scenarios with varying number of eRRHs and SHs.}
    \label{Fig6}
\end{figure}

Figure~\ref{Fig5} demonstrates the cache hit rate of an SH-aided F-RAN, including four eRRHs and various numbers of SHs in different $\gamma$ values. We can see how supplementing SHs to the network increases the Cache hit rate for all values of~$\gamma$. A higher cache hit rate brings about serving more users from the cache resources, thereby lowering fronthaul loads and improving delivery delays due to the elimination of back-and-forth communications to the MBS. The figure depicts that the efficiency of the SHs grows larger for higher values of~$\gamma$ where the popularity of the file segments changes sharper from the most to the least probable file segments. Nevertheless, the inclusion of SHs to F-RANs still remains useful for lower values of $\gamma$ as well, and it improves the load on the eRRHs by serving some of the users based on the simulation results, leading to a better quality of service and more users served in each time subframe.

Figure \ref{Fig6} represents the average delivery delay of the SH-aided networks with different numbers of eRRHs. In each subfigure of Fig.~\ref{Fig6}-(a), adding more SHs to the network improves the average delivery delay, disregarding the $\gamma$ value. However, for higher values of $\gamma$, supplementing SHs to the network minimizes the delay further. That is because users' request preference is centred around the most probable file segments, and it is more probable to serve the users from the cache resources of SHs and eRRHs. In addition, we can see that SHs can compensate for the lack of eRRHs in the network. For example, the average delivery delay of having one eRRH aided by four SHs with $\gamma = 1$ is improved 143 ms. On the other hand, the delivery delay of having two eRRHs without the SHs and with $\gamma = 1$ is 191 ms, significantly higher than the delay observed with the one eRRH and four SHs configuration. This trend could be seen for the scenarios of 3 and 4 eRRHs with and without SHs, indicating that SHs can effectively substitute eRRHs. Fig.~6-(b) compares the performance of the SH-aided F-RANs in terms of average delivery delay with those of networks assisted by CE-relay, where $R$ represents the number of CE-relays. The solid and dashed lines, respectively, represent SH-aided and CE-relay-aided networks with a certain number of eRRHs, SHs and CE-relays. We can see that SHs significantly improve the delivery delay compared to CE-relays due to a couple of reasons. First, the SHs can update their cache resources faster than CE-relays, as SHs have access to more popular data thanks to their overhearing capability. Additionally, SHs eliminate the need for communications with the MBS by only relying on their own cache resources, whereas CE-relays utilize fronthaul links in multi-hop communications.


\begin{figure}
\begin{subfigure}[b]{0.48\textwidth}
        \centering
        \includegraphics[width=\columnwidth]{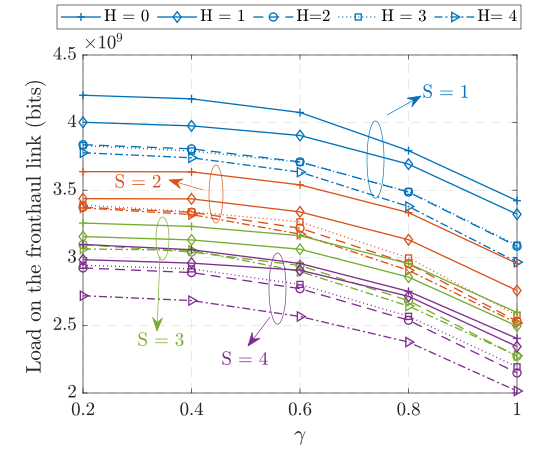}
        \caption{Average load on the fronthaul link in each transmission subframe.}
\end{subfigure}
\begin{subfigure}[b]{0.48\textwidth}
        \centering
        \includegraphics[width=\columnwidth]{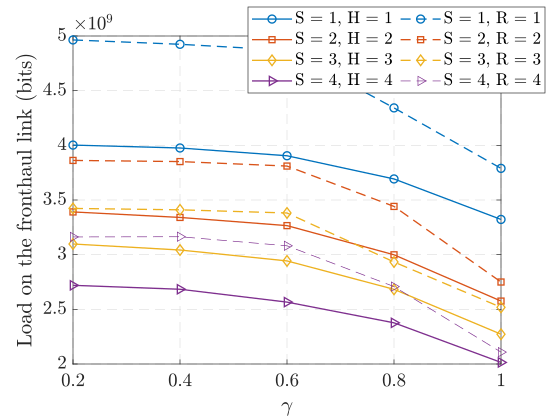}
        \caption{Comparison of the average load on the fronthaul link of the F-RANs aided with the SHs versus the CE-relays.}
\end{subfigure}
\caption{Load on the fronthaul link for different schemes varying number of eRRHs and SHs.}
\label{Fig7}
\end{figure}

In Fig.~\ref{Fig7}, the impact of using SHs on the load of the fronthaul link is shown. Fig.~\ref{Fig7}-(a) illustrates that adding up to four SHs to the network reduces the load on the fronthaul link by around $5 \times 10^8$ bits/sec. Moreover, it is noteworthy that with the addition of more SHs, the load on the fronthaul link can be further optimized, potentially surpassing the performance of the network configuration with one additional eRRH. This finding suggests that leveraging SHs strategically can effectively enhance the overall efficiency and performance of the network infrastructure.  
In addition, Fig.~7-(b) illustrates that an SH-aided F-RAN network undergoes an improved fronthaul load compared to the networks assisted by CE-relay. This is due to the fact that in CE-relay-aided networks, the cache resources of CE-relays are updated slower and some of the users are being served from the MBS utilizing the relays' capability, which leads to higher loads on the fronthaul links.

\begin{figure}
\begin{subfigure}[b]{0.5\textwidth}
        \centering
        \includegraphics[width=\columnwidth]{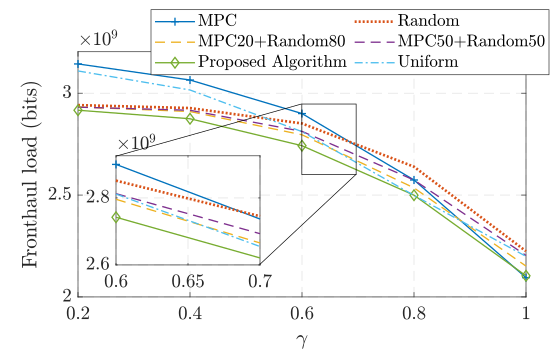}
        \caption{Comparison of the average load on the fronthaul link of SH-aided F-RANs utilizing different Caching schemes.}
\end{subfigure}
\begin{subfigure}[b]{0.5\textwidth}
        \centering
        \includegraphics[width=\columnwidth]{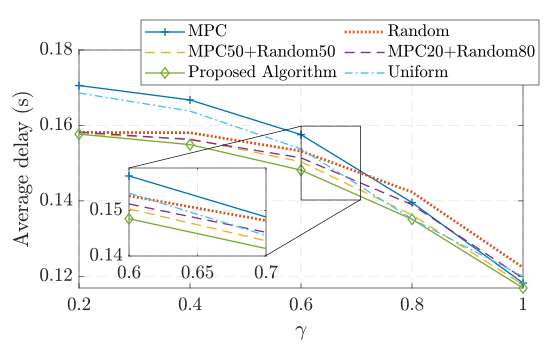}
        \caption{Comparison of the average delivery delay on the fronthaul link of SH-aided F-RANs utilizing different Caching schemes.}
\end{subfigure}
\caption{Comparison of the performance of SH-aided F-RANs utilizing different Caching schemes.}
\label{Fig8}
\end{figure}

In Fig.~\ref{Fig8}, our proposed caching algorithm is compared with five different caching schemes, including MPC, random, uniform, and two hybrid schemes combining MPC and random caching with varying ratios. The hybrid schemes are denoted as MPC20+Random80 and MPC50+Random50, where the numbers indicate the percentage allocation of caching resources to each strategy. For instance, MPC20+Random80 signifies that 20 percent of the caching resources are dedicated to caching the most probable content, while the remaining 80 percent are allocated for random caching. Similarly, MPC50+Random50 indicates an equal distribution of caching resources between MPC and random strategies, each receiving 50 percent. 
In the comparison presented in the figure, our designed caching algorithm consistently outperforms all other caching schemes across a wide range of $\gamma$ values from 0 to 1. While each alternative caching scheme exhibits superiority within specific $\gamma$ ranges, our algorithm consistently achieves the most notable improvements in both delivery delay and fronthaul load, highlighting its effectiveness in optimizing network performance and resource utilization.


\section{Conclusion} \label{SectionVI}
In this work, we proposed enhancing F-RAN resource settings cost-effectively by introducing a set of compact caching elements called SHs. These SHs function as efficient caches for popular file segments, alleviating the burden on the fronthaul link and reducing delivery delay. To optimize the network's performance in terms of average transmission delay and fronthaul link load, we devised a multistage user scheduling and caching algorithm.
Initially, we utilized an MWIS search method to assign users to eRRHs/SHs, prioritizing SHs to serve users. For caching strategies at eRRHs and SHs, we introduced two MARL algorithms to optimize their caches. Our simulations demonstrated that integrating SHs notably diminishes fronthaul link load and transmission delay, showcasing their potential in high-load and delay-sensitive wireless networks.
Furthermore, our findings suggest that SHs can replace some eRRHs to further alleviate fronthaul link load. Comparing SH-aided F-RANs with networks aided by CE-relays, we observed enhanced network performance with SH-aided F-RANs.
Finally, we benchmarked our caching scheme against existing schemes, showing its superiority regardless of file segment popularity distribution. These results underscore the effectiveness of our proposed approach in optimizing F-RAN resource settings.
\section*{Appendix\\
Proof of Theorem 1}
According to \cite[Theorm 4]{APP}, assuming $\lim_{n\to\infty} \frac{\alpha_u}{\alpha_{\pi}} = 0$, we can conclude that two conditions, $\mathbf{c1:}\lim_{n \to \infty} |u^{i,(n)}_{f,s}-\bar{\mu}_{f,s}^i| = 0$, $\forall i,f,s$; and $\mathbf{c2:}$ as $n \to \infty$, are met. Then, the equation used to update the agent's action selection probability will converge to the following ordinary differential equation
\begin{align} \label{APP1}
     \dot{\pi}^{i,(n)}_{f,s} =  \beta^{i,(n)}_{f,s}(u^{i,(n)}_{f,s}) - \pi^{i,(n)}_{f,s}.
\end{align}
Considering $\lim_{n\to\infty} \frac{\alpha_u}{\alpha_{\pi}} = 0$, the update of action selection probabilities occurs at a slower pace in comparison to the estimated utility. However, as per condition $\mathbf{c1}$, the estimated utility will converge as time progresses. By substituting the converged estimated utility value into (\ref{APP1}) and considering the fact that at the stationary point of (\ref{APP1}), $\dot{\pi}^{i,(n)}_{f,s}=0$, we can determine the value of the stationary action selection probability,
\begin{align}\label{APP2}
    \bar{\pi}^{i,(n)}_{f,s} = \beta(\bar{\mu}_{f,s}^i).
\end{align}
\begin{figure*}[hbt!] 
 \begin{align}\label{APP3}
 \{\bar{\pi}^{(n),1}_{f,s},\bar{\pi}^{(n),2}_{f,s}\} =  \arg\max_{\pi^{1}_{f,s} \geq 0, \pi^{2}_{f,s} \geq 0} \sum_{i = 1}^{2} \pi^{i}_{f,s} \left(\bar{\mu}_{f,s}^i + \frac{1}{\lambda_p} \ln (\frac{1}{\pi^{i}_{f,s}}) \right),~~
 \text{s.t.} ~~\sum_{i = 1}^{2} \pi^{i}_{f,s} = 1
 \end{align}
\hrulefill
\end{figure*}
We proceed to explain that (\ref{APP2}) gives the resulting probability for selecting actions among the agents. Reference \cite[eq. (2)]{APP2} indicates that the most favourable action selection probability is achieved by maximizing the agents' expected reward along with the entropy function. Therefore, we derive (\ref{APP3}) for each agent ($s,f$), $f \in \{1,2,\dots, F\}$ and $s \in \{\{1,2,\dots, S\}$, which is shown at the top of the next page. The value $1/\lambda_p$ in (\ref{APP3}) determines the relative importance of maximizing the expected reward and entropy functions. By using the Lagrangian optimization method, we can easily calculate the optimal values
\begin{align}
     \bar{\pi}^{i,(n)}_{f,s} = \frac{\exp(\lambda_p \bar{\mu}_{f,s}^i)}{\sum_{i=1}^2 \exp(\lambda_p \bar{\mu}_{f,s}^i)}  \triangleq \beta(\bar{\mu}_{f,s}^i). 
\end{align}
It is apparent that when $n \to \infty$, the likelihood of agents selecting an action becomes consistent, and this consistent likelihood aligns with the ideal action selection probability for the agents. Hence, as the number of iterations increases, Algorithm \ref{Algorithm2} converges.
\bibliographystyle{IEEEtran}
\bibliography{references}
\end{document}